# Long-lived refractive index changes induced by femtosecond ionization in gas-filled single-ring photonic crystal fibers

Johannes R. Koehler[*], Felix Köttig, Barbara M. Trabold, Francesco Tani, and Philip St.J. Russell
*Max Planck Institute for the Science of Light, Staudtstr. 2, 91058 Erlangen, Germany*
*[johannes.koehler@mpl.mpg.de](mailto:johannes.koehler@mpl.mpg.de)

We investigate refractive index changes caused by femtosecond photoionization in a gas-filled hollow-core photonic crystal fiber. Using spatially-resolved interferometric side-probing, we find that these changes live for tens of microseconds after the photoionization event — eight orders of magnitude longer than the pulse duration. Oscillations in the megahertz frequency range are simultaneously observed, caused by mechanical vibrations of the thin-walled capillaries surrounding the hollow core. These two non-local effects can affect the propagation of a second pulse that arrives within their lifetime, which works out to repetition rates of tens of kilohertz. Filling the fiber with an atomically lighter gas significantly reduces ionization, lessening the strength of the refractive index changes. The results will be important for understanding the dynamics of gas-based fiber systems operating at high intensities and high repetition rates, when temporally non-local interactions between successive laser pulses become relevant.

## I. INTRODUCTION

The interaction between a femtosecond laser pulse and a gas is governed primarily by the fast evolution of the material polarizability and is usually treated on a single-shot basis, even when the intensities are sufficient for photoionization. Although every ionization event is followed by long-lived thermal and hydrodynamic effects (on ns to µs timescales), such effects are irrelevant on the fs timescale of a single driving pulse and are therefore often disregarded. Furthermore, conventional laser sources, with pulse energies high enough (several mJ range) to photo-ionize a gas, operate at relatively low repetition rates (typically below 1 kHz) – too slow to observe plasma-related inter-pulse interactions. Ionization-induced effects in the µsec range are typically investigated with a sequence of two pulses, the second pulse acting as a probe. With such a pump-probe scheme it is possible, for example, to investigate plasma waveguides [1,2] and thermal waveguides in air [3].

In systems capable of reaching the strong-field regime with pulse energies in the several µJ range, it is possible to operate at MHz repetition rates without a great increase in the average power compared to kHz systems. Under these circumstances the behavior is found to deviate dramatically from the single-shot case.

Two examples of such systems are femtosecond enhancement cavities and gas-filled hollow-core photonic crystal fibers (PCFs). At high pump-laser repetition rates, any long-lived dynamics will cause temporally non-local pulse-to-pulse interactions, making synthesized pump-probe pulse sequences unnecessary. In femtosecond enhancement cavities operating at tens-of-MHz repetition rates, this caused the build-up of a constant background plasma density which impairs phase matching for high-harmonic generation, thereby reducing its efficiency [4]. Similarly, recent experiments on pulse compression and UV generation in gas-filled hollow-core PCF have uncovered a significant dependence of the nonlinear dynamics on repetition rate, even at 100 kHz [5]. These results illustrate the importance of long-lived photoionization effects, the understanding of which will be crucial for scaling to even higher repetition rates, in both free-space set-ups and gas-filled PCFs.

Here we report a detailed experimental study of refractive index changes in a gas-filled hollow-core PCF that occur after femtosecond photoionization. An interferometer is used to side-probe these changes, permitting their temporal and spatial evolution to be monitored over lifetimes of several tens of µs. At the same time, plasma-induced sound waves emerge after free-carrier recombination [6,7], driving mechanical resonances in the thin-walled capillaries surrounding the PCF core. Our results imply that deviations from single-shot behavior will become apparent even at repetition rates as low as tens of kHz.

## II. EXPERIMENTAL SETUP

A typical gas-filled hollow-core PCF provides anomalous dispersion at gas pressures up to a few bar, so that a pump pulse with soliton order $N > 1$ will undergo self-compression as it propagates along the fiber. In this way a pulse with a few µJ of energy and a duration of few tens of fs can be compressed down to (or even below) a single optical cycle. At such a temporal focus the peak intensity is high enough to ionize the gas [8].

In the experiments, a 26-cm-long single-ring hollow-core PCF (SR-PCF) was placed between two gas cells, each sealed by a 3-mm-thick uncoated silica window (Fig. 1(a)). A scanning electron micrograph of the SR-PCF microstructure is shown in Fig. 1(b). Six silica capillaries (inner diameter ~42 µm and wall thickness ~420 nm) are mounted inside a thick-walled glass capillary. The central hollow core is ~57 µm in diameter. The normal dispersion introduced by 2 bar of argon is sufficient to counterbalance the anomalous dispersion of the hollow core, producing zero dispersion at ~550 nm and anomalous dispersion at longer wavelengths. Pulses with duration of 39 fs and central wavelength 805 nm were delivered at 1 kHz repetition rate by a Ti:sapphire laser amplifier and launched into the fundamental core mode using an achromatic lens with 25 cm focal length. Pulses of energy 19 µJ were used to generate a plasma at the temporal focus.

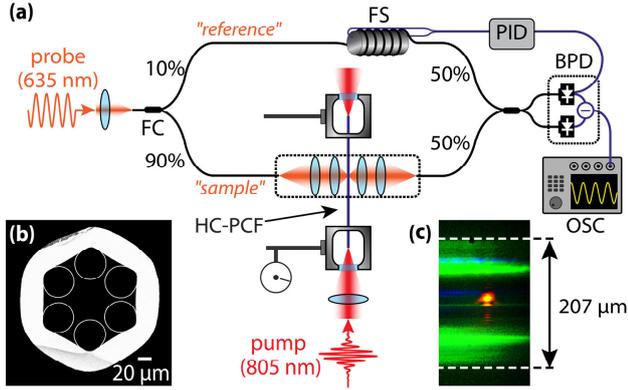

**Fig. 1.** (a) Experimental setup. Pump pulses (red) propagate along a 26-cm-long SR-PCF filled with 2 bar of argon. The side-probing Mach-Zehnder interferometer comprises: FC, fiber coupler; FS, fiber stretcher; BPD, balanced photo detector; OSC, oscilloscope; PID, stabilization controller. Imaging the fiber endface while mounting ensured correct orientation in the gas cells. (b) Scanning electron micrograph of the SR-PCF microstructure. Six capillaries (wall thickness 420 nm, inner diameter 42 µm) are mounted within a thick-walled capillary. The core diameter is ~57 µm and the gaps between the capillaries are ~7 µm wide. (c) Optical micrograph of the fiber taken from the side (green light: external illumination, orange spot: focused probe beam, dashed lines: outer cladding edges).

To detect transient refractive index changes following photoionization, we employed a fiber-based Mach-Zehnder interferometer similar to the one reported in [9]. The light source was a stabilized 15 mW diode laser at 635 nm, with < 28 pm linewidth. In the "sample" arm of the interferometer (perpendicular to the SR-PCF axis), the probe light was focused into the fiber core through the ~7-µm-wide gap between two capillaries (Fig. 1(c)). The transmitted light was then recollimated and collected by a single-mode fiber. The focusing and recollimating lenses were chosen so that the Rayleigh length (~200 µm) matched the outer diameter of the SR-PCF while yielding a beam-waist of ~7 µm. Although under these circumstances ~96% of the probe light passed the gap, the remaining fraction unavoidably overlapped with the capillaries which, when deflected, would cause an additional phase change. All the optical components in the free-space section of the sample arm were mounted on a computer-controlled translation stage, allowing the probing position to be scanned along the 12.5-cm-long section of the SR-PCF between the gas cells. The phase-modulated probe light in the "sample" arm was combined at a 2×2 fiber coupler with the unmodulated "reference" light. The resulting intensity modulation was measured at both output ports using an amplified balanced photo-detector, which suppresses noise due to laser intensity fluctuations.

A piezo-based fiber stretcher in the "reference" arm, controlled by a feedback loop, allowed the working point of the interferometer to be stabilized at quadrature, thus maximizing its sensitivity. An oscilloscope triggered to the Ti:sapphire output was used to acquire 1-ms-long temporal traces of the signal at the balanced photo-detector at a 200 MHz sampling rate. To reduce effects of phase noise in the probe laser and interferometric drift, the traces were averaged sixty times before saving and an additional ten times in post processing.

### III. RESULTS

Figure 2(a) plots temporal traces of the probe phase change at two positions: 12.2 cm (①) and at 14.2 cm (②) from the input fiber end. The standard deviation of the signal for $t < 0$ (the pulse reaches the probing position at $t = 0$) is ~0.2 mrad, indicating that it is possible to resolve phase changes as small as ~0.4 mrad with a signal-to-noise ratio of 3 dB.

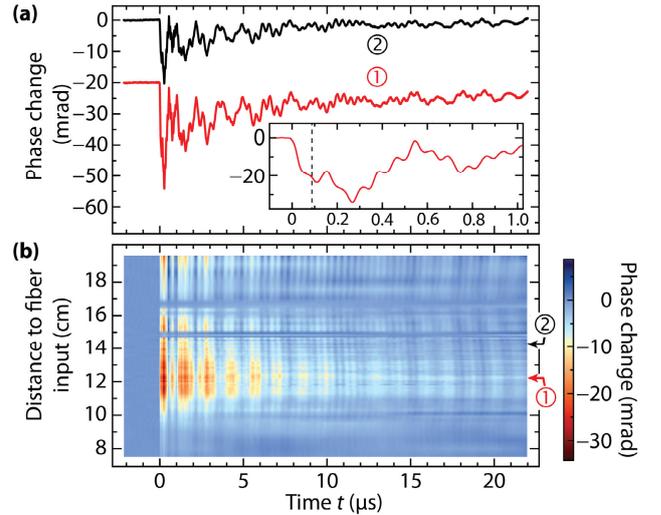

**Fig. 2.** (a) Temporal response of the probe phase at the ~12.2 cm (①) and the ~14.2 cm (②) positions along the fiber. For clarity the traces are offset by 20 mrad. Inset: zoom-in of trace ①. The dashed line marks the estimated onset of periodic modulation. (b) Full spatio-temporal evolution of the probe phase along the fiber.

At $t = 0$, ionization causes a rapid drop in refractive index directly underneath the envelope of the ionizing pulse. The index change in the presence of free electrons is given by $\Delta n_g \approx -N_e e^2 / (2 n_g \omega^2 \varepsilon_0 m_e)$, where $N_e$ is the free electron density, $e$ the electronic charge, $n_g$ the index of the neutral gas, $\omega$ the angular frequency of the light, $\varepsilon_0$ the vacuum permittivity and $m_e$ the effective electron mass. The measured probe phase (inset of Fig. 2(a)) shows a step-like change, convoluted with the much slower response of the balanced photo-detector (the rise time from 20% to 80% is ~10 ns). The plateau in probe phase at $t \approx 45$ ns can be attributed to the interplay of two effects. Firstly, electron recombination within ~10–30 ns [10–12], resulting in a gradual increase in refractive index (ionized argon has the longest recombination time of all the noble gases). And secondly, heating due to recombination, which induces a local overpressure proportional to $N_e$ [13]. This triggers an outward-propagating pressure wave that leads to rarefaction in the core, producing a negative refractive index change [3,6,7,13] that initially grows until, after a few tens of nanoseconds, it approaches a quasi-steady state characterized by pressure equilibrium and an elevated temperature within the former plasma region [13].

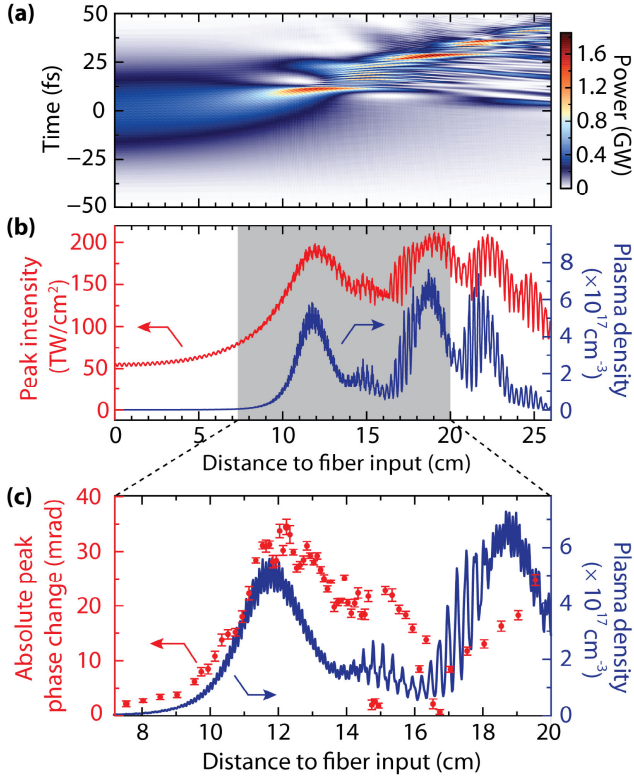

**Fig. 3.** (a) Simulated temporal evolution of a Gaussian pulse propagating in the gas-filled SR-PCF. (b) Simulated peak intensity (red) and on-axis plasma density (blue) along the fiber. Gray shading indicates the range accessible to side-probing. (c) Absolute value of the measured peak phase change (red dots) and simulated on-axis free electron density (blue line) along the fiber. The data points (dots) were averaged ten times, and the error bars indicating the standard deviation.

Propagating at the speed of sound in neutral argon (323 m/s [14]), the pressure wave requires ~90 ns to reach the capillaries (dashed line in Fig. 2(a)). After it impinges, several mechanical modes of the capillaries are impulsively excited and a beat pattern appears in the probed temporal traces. The inset in Fig. 2(a) shows for example periods of 76 ns and 478 ns, corresponding to frequencies of 2.1 and 13.2 MHz. While mechanical damping causes the beating to gradually fade away, the average phase change measured by the probe follows the slow decay of the density depression by thermal diffusion and drops within ~10 µs to half of its maximum value.

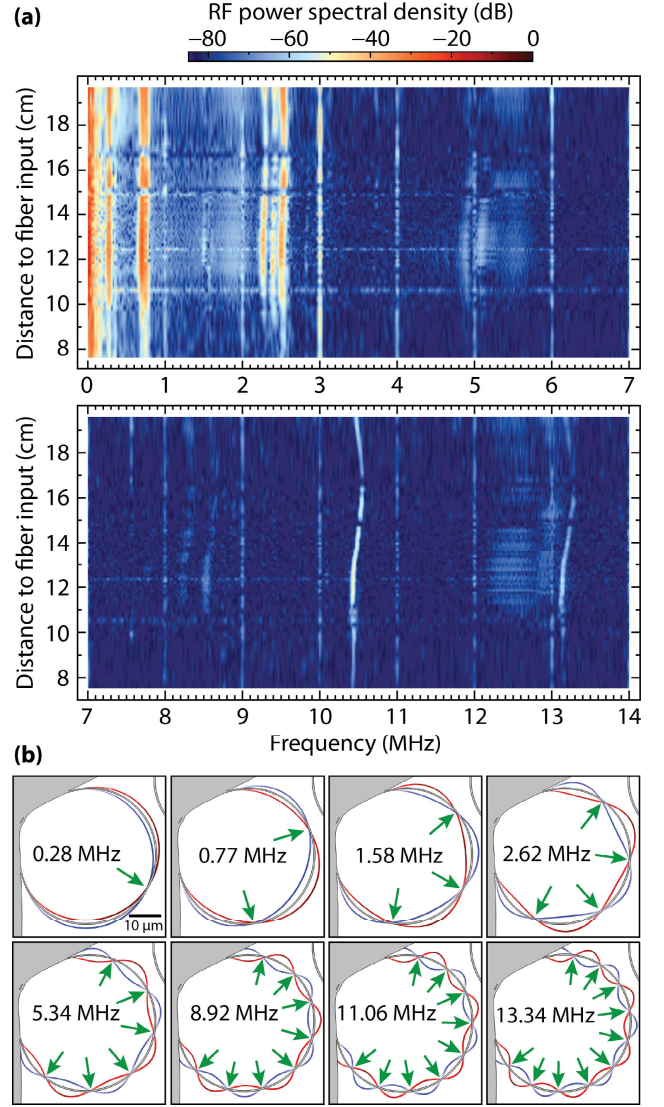

**Fig. 4.** (a) Frequency spectrum of the transient phase change shown in Fig. 2(b), plotted versus distance from the fiber input. Peaks (or clusters) are seen at 0.3, 0.7, 2.4, 3.7, 5, 8.4, 10.4 and 13.2 MHz. (b) Deflection of the left upper capillary (not to scale) by eight flexural resonances with different number of nodes (indicated by green arrows).

Integrating the phase along the probe direction, a simple diffusion model based on a Gaussian temperature distribution [13] predicts a timescale of $\sim 3r_{pl}^2/(4\alpha) = 4$ μs where $r_{pl}$ is the radius of the plasma and $\alpha \approx 0.1$ cm$^2$/s is the thermal diffusivity in argon. Scanning along the fiber reveals that, while the temporal dependence of probe phase maintains its shape, its overall amplitude varies with position (Fig. 2(b)).

The evolution of the pump pulse was numerically modeled using a unidirectional field equation [15], based on an analytic model for the fiber dispersion [16] that included the effect of anti-crossings between the core mode and capillary-wall resonances. Figure 3(a) shows the calculated temporal evolution of an unchirped 39 fs (FWHM) Gaussian pulse with 16 μJ energy (slightly lower than in the experiment, so as to account for imperfect launching into the fundamental mode and fiber loss). As it propagates, the pulse undergoes soliton self-compression, reaching its shortest duration at ~12 cm from the fiber input face, with a peak intensity as high as $\sim 2 \times 10^{14}$ W/cm$^2$ (Fig. 3(b)). After the point of maximum compression, soliton fission occurs, followed by several additional cycles of self-compression, with intensity maxima at ~19 cm and ~22 cm. Figure 3(b) plots the peak intensity along the fiber together with the free electron density ($N_e$), calculated using Ammosov-Delone-Krainov (ADK) ionization rates [17]. The simulation predicts a considerable level of ionization, with pronounced peaks in $N_e$ (up to $\sim 7 \times 10^{17}$ cm$^{-3}$) at the three intensity maxima, while the electron density over the initial ~10 cm, before the pulse has fully compressed, is two orders of magnitude smaller (less than $\sim 10^{16}$ cm$^{-3}$). Figure 3(c) plots a zoom-in of $N_e$ along the experimentally probed fiber region, along with the largest detected probe phase change at each position (found at time $t \approx 260$ ns, see Fig. 2(a)). Although the two curves agree qualitatively well, the measured phase change is roughly twice as large as expected from the simulated plasma density, which we attribute to capillary resonances. The strong deviations at ~14.9 cm were insensitive to any realignment of the probe beam, and are likely due to imperfections on the outer fiber surface.

Figure 4(a) plots the temporal traces from Fig. 2(b) in the Fourier domain as a function of radio frequency (RF) and position along the fiber. The signals at integer multiples of 1 MHz, also present in the absence of any pump light, result from phase-noise peaks in the probe laser light. In the region where $N_e$ is significant, i.e, after the initial ~10 cm (Fig. 3(b)), clusters of peaks emerge at frequencies around ~0.3, 0.7, 2.4, 8.4, 10.4 and 13.2 MHz and up to 50 dB above noise level. These are associated with various flexural resonances in the capillaries, characterized by their number $m$ of nodes around the azimuth (Fig. 4(b)). The capillaries are not perfectly identical, so that their respective eigenfrequencies for the same value of $m$ are different. Predictions of their frequency ranges by finite element modeling (FEM), based on the actual fiber structure, are listed in Tab. 1 and found to be consistently slightly higher than the experimental values. We attribute this to mechanical damping, not accounted for in the FEM, and to inaccuracies in digitizing the fiber structure. We note that the mechanical resonant frequencies also vary slightly over the fiber length (Fig. 4(a)), with larger deviations for higher mode orders, indicating that the capillaries are not perfectly uniform along the fiber.

| $m$ | Frequency (MHz) | $m$ | Frequency (MHz) |
|---|---|---|---|
| 1 | 0.3–0.4 | 6 | 5.3–6.9 |
| 2 | 0.8–1.0 | 7 | 7.0–9.1 |
| 3 | 1.6–2.0 | 8 | 8.9–11.6 |
| 4 | 2.6–3.4 | 9 | 11.1–14.3 |
| 5 | 3.9–5.0 | 10 | 13.3–17.3 |

**Tab. 1.** Modeled eigenfrequencies of flexural vibrations in the capillaries, as a function of the number $m$ of nodes around their azimuth.

Weak signals are also seen at 1.5 and 3.7 MHz, with linewidths that are identical to those of the stronger signals. We attribute these to $m = 3$ and 5 modes (Tab. 1). Their much lower signal-to-noise ratios (less than 20 dB) can be explained by the interplay of two effects. Firstly, although excitation of these modes by a plasma-induced sound wave emanating from core center would be impossible in a perfectly symmetric fiber structure (because the point of impact on the capillaries would coincide with a node in the mechanical deflection profile (see Fig. 4(b)), deviations from perfect six-fold symmetry permit this in the real fiber. Secondly, in strong contrast to $m = 1$, the mode shapes for $m = 3$ and 5 cause only slight changes in the amount of silica glass in the beam path during one vibrational cycle, resulting in a smaller probe phase variation.

In regions along the fiber with large $N_e$, an extremely weak signal (only ~7 dB above noise level) appears at ~5.5 MHz, with a linewidth of ~300 kHz (10 times larger than that of the capillary resonances). This is caused by the sound wave oscillating back and forth across the fiber core, forming an acoustic cavity. Measurements with a counter-propagating probe laser, to be reported elsewhere, confirm this.

To explore the effect of reduced plasma density, we repeated the same experiment with neon, a gas with a much higher ionization potential. At 20 bar of neon the dispersion and nonlinearity are comparable to the values for argon at 2 bar, resulting in almost identical pulse-propagation dynamics and optical field distributions. Numerical simulations predict a ~9 times smaller maximum plasma density and therefore a weaker pressure wave than in argon. In the experiment, we indeed observed a much smaller signal from the vibrating capillaries, only slightly above the detection limit, confirming that a pressure wave is the dominant driving mechanism and ruling out other effects unrelated to ionization (in particular, optomechanical forces that could arise from differences in the optical fields on opposite sides of the capillary walls).

## IV. CONCLUSIONS

The temporal and spatial characteristics of plasma-induced refractive index changes due to self-compressed laser pulses propagating in the gas-filled core of a SR-PCF can be conveniently monitored by interferometric side-probing. Measurements reveal a typical lifetime of ~20 µs, which is related predominantly to the dynamics of recombination-driven heating in the gas. MHz mechanical resonances in the thin-walled capillaries surrounding the SR-PCF core are simultaneously excited by a plasma-driven acoustic wave.

These observations have two important consequences. Firstly, non-local interactions between successive pump pulses could occur at repetition rates as low as 50 kHz. Secondly, the relatively large mechanical linewidths mean that the laser repetition rate could accidentally lie within the Lorentzian lineshape of a capillary resonance and drive it synchronously, enhancing the vibrational amplitude. This could be prevented by employing SR-PCFs with smaller capillary diameters whose resonant frequencies are higher than the pump repetition rate.

The impact of these long-lived refractive index changes can be minimized by employing atomically lighter gases with higher ionization potential. For example, replacing 2 bar of argon with 20 bar of neon reduced the free electron density and the initial vibrational amplitude by a factor of ~9. The damping rate of the mechanical vibrations scales with $(m_a)^{0.5}p$, where $m_a$ is the atomic mass of the gas and $p$ its pressure [18]. As a result, an additional advantage of using higher pressure neon is a ~7 times larger gas damping rate; this is because of the higher pressure needed to achieve a similar dispersion landscape.

Besides improving the fundamental understanding of hollow core PCF-based laser systems at high pulse energies, it will be essential to take the results in this letter into account when scaling to high repetition rates.